\documentstyle[twoside]{siamltex}

\title{Periodic generation and propagation of traveling
fronts in dc voltage biased semiconductor superlattices\thanks{Received
by the editors of SIAM J. Appl. Math. on  11 July, 1995. Manuscript number 
028888-1.}}
\renewcommand{\theequation}{\arabic{section}.\arabic{equation}}

\author{Luis L. BONILLA\footnotemark[2] \and Manuel KINDELAN\footnotemark[2]
 \and Miguel MOSCOSO\footnotemark[2] \and Stephanos VENAKIDES\footnotemark[3]}   

\begin{document}

\maketitle

\renewcommand{\thefootnote}{\fnsymbol{footnote}}

\footnotetext[2]{Escuela Polit{\'e}cnica Superior, Universidad Carlos III 
de Madrid, Butarque 15, 28911 Legan{\'e}s, Spain.
The research of this author was supported by  DGICYT grants PB92-0248 
and PB94-0375, by North Atlantic Treaty Organization
traveling grant CRG-900284, and by EEC Human Capital 
and Mobility Program grant ERBCHRXCT930413.}

\footnotetext[3]{Department of Mathematics, Duke University,
Durham, NC 27708, USA. The research of this author was supported by 
the Army Research Office grant DAAH04-93-G-0011, National Science 
Foundation grant DMS-91-03386, and by North Atlantic Treaty 
Organization traveling grant CRG-900284.}

\renewcommand{\thefootnote}{\arabic{footnote}}

\begin{abstract}
The continuum limit of a recently-proposed model for charge transport
in resonant-tunneling semiconductor superlattices is analyzed. It is
described by a nonlinear hyperbolic integrodifferential equation on a
one-dimensional spatial support, supplemented
by shock and entropy conditions. For appropriate parameter values, a
time-periodic solution is found in numerical simulations of the model. 
An asymptotic theory shows that the time-periodic solution is due to
recycling and motion of shock waves representing domain walls connecting
regions of the superlattice where the electric field is almost uniform. 
\end{abstract}


35G25, 35M05, 35B25.

\pagestyle{myheadings}
\thispagestyle{plain}
\markboth{L. L. BONILLA, M. KINDELAN, M. MOSCOSO AND S. VENAKIDES}{Traveling
fronts in semiconductor superlattices}




%
\setcounter{equation}{0}
\section{Introduction}
\label{sec-introduction}

Self-sustained oscillations of the current have been observed in
recent experiments on n-doped weakly-coupled semiconductor superlattices
(SLs) under $dc$-voltage bias along the growth direction \cite{grahnprl}. The
main charge transport mechanism is quantum resonant tunneling through
the SL \cite{esaki,kazar,grahn90}. Similar but 
damped oscillations have been reported earlier in undoped superlattices subject 
to laser photoexcitation \cite{merlin}. In both cases a phenomenological 
discrete drift model proposed by one of the authors \cite{bonillaSL} explains 
the oscillations in terms of the formation and dynamics of electric field 
domains, i.e. regions in space in which the strength of the electric field 
is nearly uniform. These phenomena are thus examples of time-dependent 
pattern formation in resonant-tunneling SL, a scarcely explored area of 
semiconductor physics.

We shall study asymptotically and numerically 
the self-sustained oscillatory solutions of the continuum limit  of the model 
that applies when the number of quantum wells in the SL is large or equivalently 
when the SL is long. The continuum limit of the model, 
to be derived below, is given by the equations (see \cite{ICPF94})
\begin{eqnarray}
\frac{\partial E}{\partial t} + v(E)\, \left( 1 + \frac{\partial E}{\partial 
x}\right) &=& I,  \label{1-1}  \\
\frac{1}{L}\,\int_{0}^{L} E(x,t)\, dx &=& \phi.           \label{1-2} 
\end{eqnarray}
The unknowns in these equations are the current density $I(t)$ and the  
electric field $E(x,t)$ on a one-dimensional SL $0<x<L$. The velocity $v(E)$, 
shown in the inset of Figure 1a, is a positive
function for $E>0$ having two peaks at $E=1\, $ [$v(1)=1$] and $E=E_M >1$ 
[$v(E_M)=v_M > 1$] separated by a minimum at $E=E_m \in (1,E_M)$ [$v(E_m)=v_m < 
1$]. The constant $\phi$ is a control parameter proportional to the $dc$ 
voltage bias. The self-sustained oscillations are stable time-periodic solutions
that appear for $1<\phi<E_m$; see the oscillation of the current density in
the inset of Figure 1b.
Equations  (\ref{1-1}) and (\ref{1-2}) are to be solved with the 
boundary condition \cite{ICPF94}:
\begin{eqnarray}
\frac{\partial E(0,t)}{\partial x} &=& c > 0. \label{1-3}
\end{eqnarray}
Eq.\ (\ref{1-1}) is hyperbolic, so its solutions will in general develop
shock waves. We supplement the model with the shock condition \cite{ICPF94}:
\begin{eqnarray}
\int_{E_{-}}^{E_{+}} \left( \frac{1}{v(E)} - \frac{1}{V(E_{+},E_{-})} \right) 
\, dE = 0,	\label{1-4}
\end{eqnarray}
and the entropy condition \cite{ICPF94}
\begin{eqnarray}
v(E_- )\geq V(E_{+},E_{-})\geq v(E_+ ) . \label{1-5}
\end{eqnarray}
Here $V(E_+,E_-)$ is the velocity of a shock wave located at $x=X(t)$ and such 
that $E(X-,t) = E_-$ and $E(X+,t) = E_+$.

Physically, Eq. (\ref{1-2}) expresses the fact that the voltage drop across the 
semiconductor length is kept constant at the value $\phi\L$.  Eq. 
(\ref{1-1}) can be derived directly from Ampere's Law  
\begin{eqnarray}
\frac{\partial E}{\partial t} + v(E)\, n &=& I,  \label{ampcont}  
\end{eqnarray}
by substituting in it the Poisson equation
\begin{eqnarray}
\frac{\partial E}{\partial x} = n -1,
\end{eqnarray}
 where the one in the right hand side represents the normalized density 
of the dopant. 
The electron current $v(E)\, n$ is essentially given by the probability of
tunneling from one quantum well of the SL to the next one, and is assumed to be
proportional to the electron density in the quantum well. Our derivation of 
(\ref{1-1}) homogenizes over the SL structure. The quantum effects due to the 
SL structure that dominates the phenomenon are modeled in the shape of the 
function $v(E)$ (the electron velocity) and in particular in its having two 
peaks as shown in the inset of Fig.\ 1a above. Peaks occur at certain values of 
the field as a result of energy level alignment in adjacent quantum wells that 
leads to enhanced (or resonant) tunelling and hence to enhanced transmission.

The shock condition (\ref{1-4}) arises in the passage from the discrete 
model to the continuum model. It has the geometric interpretation of an equal 
area rule as one observes in Figure 2. In essence the shock condition provides a 
relation between the field $E_+$ at the front of the shock wave, the field 
$E_-$ at the back of the shock and the shock speed $V$. The domain of this 
relation is restricted by the requirement that the entropy condition be 
satisfied. The entropy condition (\ref{1-5}) is the usual one, saying that 
the particles entering the shock move faster than the shock itself, which in 
turn moves faster than the particles it encounters. The entropy condition also
ensures that the electron density inside the shock wave (see below for the
meaning of a region of the SL inside of the shock wave) is positive. 

Our calculation is organized as follows: In 
 Section \ref{sec-qualitative} we discuss the oscillations from a 
qualitative viewpoint and in the further sections we do the analysis of
one complete cycle of the oscillations. 
We start by analyzing the
Shock Kinematics, 
in Section \ref{sec-kinematics},
and explain analytically the numerically observed monopole 
(shock plus tail) structure. In section \ref{sec-asymptotics} we introduce our 
asymptotic scaling, derive the outer and inner solutions and put the pieces 
together working and matching in sequence the different regions which 
describe a whole oscillation cycle. 
 In sections \ref{sec-birth} and \ref{sec-death} we analyze in detail 
the process of creation and disappearance of a monopole, and in section 
\ref{sec-discussion} we  include some final remarks. Finally
Appendix \ref{sec-discrete} contains a derivation of the continuum  
limit of the discrete SL model.

\setcounter{equation}{0}
\section{Qualitative Discussion of Current Oscillations}
\label{sec-qualitative}

We study the asymptotic limit in which the length of the SL sample is large.
We have derived the time-periodic solution of the above model asymptotically 
and numerically. We describe it in Figure 1, in which the
numerical solution of equations (\ref{1-1})-(\ref{1-3}) is represented.
This numerical solution has been obtained by the method of lines, using
backward differences in space, a fourth order Runge-Kutta method in time,
and a large number of grid points (5000)
in order to have sufficient precision
in the numerical approximation. Notice that this method is, in fact, equivalent
to solving the discrete model (\ref{1})-(\ref{4}) as an initial value problem.
We have also solved equations (\ref{1-1})-(\ref{1-3}) using a fully
implicit finite difference scheme and including an artificial
diffusion term proportional to $v(E)$. The key element of this method is 
a modification of the method proposed in \cite{sserna} which distributes
the grid points according to a monitor function sensitive to the local
values of the slope of the field distribution, and tuned to track the 
travelling internal layers. The results obtained with both methods
agree.

One observes that at all 
times the field is essentially uniform over two or three spatial intervals 
called domains. The latter are separated by one or respectively two thin 
internal layers (IL) in which the field value 
experiences a positive jump. The IL(s) are monotonically
increasing in $x$. 
To follow a complete cycle, we start with two domains separated by a 
single IL. Before the 
IL has reached the end of the sample $x=L$, a second IL starts forming at an 
interior point to the left of the first one, and also moves to the right. 
The field now displays two ILs separating three domains. As time still 
increases, the IL on the right disappears either because it reaches the right 
end $x=L$ of the sample (for $1\leq\phi < \phi_d$), or because the jump between 
the field values to its left and to its right decreases to zero (for $\phi > 
\phi_d$, where $1<\phi_d<E_m$). This leaves a field profile that has only two domains. The cycle then 
repeats itself. Throughout the cycle, the uniform value of the field in each 
domain varies continuously and periodically with time. At each time, the field 
is an increasing function of the spatial variable. See Figure 1.

In each IL, we find that a shock connects the left or right 
value of the field with an intermediate field  value. The latter, in turn, 
connects to the value on the other side of the IL by an exponential tail  
that moves rigidly with the shock. The characteristics are parallel to the 
shock line on the side of the tail and end into the shock line transversely on 
the other side. We call the combination af a shock and a tail a ``monopole 
wavefront''. In this language, an internal layer (IL) is a monopole wavefront 
that consists of a shock and a right tail or a shock and a left tail. As we 
explain in section \ref{sec-kinematics}, the structure of a shock with a left 
AND right tail does not appear.

As the field displays the time-periodic  spatial pattern 
formation described above, the current performs a simple oscillation as 
seen in the inset of Figure 1b. Roughly speaking, it increases with time 
as long as there 
are only two field domains and decreases when there are three field domains. We 
can describe how the field values on the domains depend on the current without 
reference to the position of the shocks in the following way. In the asymptotic 
scaling of interest [see Eqs.\ (\ref{3-1}) to (\ref{3-4}) below], 
in which the length of the semiconductor sample and the speed of the shocks are 
of order O(1), the quasistatic approximation $I=v(E)$ is valid as long as the 
current value is between the local maximum and the local minimum of the $v$ 
curve (see the
inset of Fig.\ 1a) and not too close to these extreme values. The three 
roots of the equation above are denoted  by $E^{(1)}(I) < E^{(2)}(I) < 
E^{(3)}(I)$. As the current increases [region I, times (1) and (2) in  
Figures 1] taking values in the interval $(v_m,1)$, there are only two domains
separated by an IL. The field on the left domain takes the value 
$E^{(1)}(I)$, while on the right domain it takes the value $ E^{(3)}(I)$.
The shock condition and the entropy condition, plus the requirement that 
monopoles can have only a left OR (exclusive) right tail allow us to determine 
uniquely the shock and tail characteristics as functions of the current.
After the current reaches its maximal value, a value that, as we will see, is a 
little higher than the peak of the $v$ curve, a new shock is created (region II,
time (3) in Figure 1). As the current value now decreases (Region III,
time (4) in Figure 1), there are three field domains. The field value on 
the left domain is $E^{(1)}(I)$, the value on the middle domain is $E^{(2)}(I)$,
and on the right domain it is $ E^{(3)}(I)$. Again, using the shock and entropy 
conditions, we can determine uniquely the characteristics of the two shocks and 
tails. This determination is shown graphically in the five pictures in Figure 3 
that describes the whole cycle. We are forced to graph $1/v$  instead of the 
more natural $v$ in this figure, because $v$ appears in the denominator of the 
shock condition. Thus, for example, instead of picturing the current increasing 
we think of $1/I$ decreasing. The monopole on the right disappears (Region IV; 
not shown in the figures) as the current is close to its minimal value, a value 
that is equal or slightly higher than the minimum of the $v$ curve, depending 
on the bias. 
  
The position of the IL in region II is directly
determined from the voltage bias condition as a function of $I$. 
Since the shock position and speed are known functions of the current, we 
easily obtain an ode for the current as a function of time from which 
the time-dependence of the current and of the shock position may be derived. 
A similar but more involved calculation holds in region III, where the current 
decreases taking values in the interval $(v_m,1)$.  We have three 
relations: a formula for each of the two shock speeds as functions of the 
current and the voltage bias condition. The latter 
is a linear relation between the positions of the two shocks in  which the 
coefficients are known functions of the current. We have three unknowns, 
the two IL positions and the current. We can determine 
the unknowns as functions of time given appropriate initial conditions. 
We provide the required initial conditions by (a) calculating the time and 
position of shock formation and (b) matching with the field in region II, in 
which the current is near its peak value. 
In  regions II and IV the quasistationary approximation is not always valid. We 
use fast scale variables to describe the birth and death of a shock shock in 
detail and match them to the regions I and III.

	We can represent the evolution of the shock wave without 
consideration to its spatial trajectory on the SL by means of the following
diagram. We define
the function $E_+ = {\cal F}(E_-,u)$ (where $u=1/V$) by solving the equal area 
rule (\ref{1-4}) for $E_+$. This function yields the field in front of the 
shock wave, given appropriate values of the shock speed and of the field 
at the back of the shock. The {\em domain} of this function consists 
of all the points $(E_-,1/V) $ for which there is an $E_+$ such  
that the equal area rule and the entropy inequalities are true. 
Thus a point in the domain characterizes a shock completely.
In Fig.\ 4 we have depicted the domain of ${\cal F}(E_-,u)$
and the complete life-cycle of a shock wave in it. As one can see in the 
figure and easily derive,
the domain of ${\cal F}(E_-,u)$ is bounded below by the graph of the function 
$1/v(E)$. It is bounded above by a line on which $E_+$ reaches the maximal 
value obtained by the equal area rule that does not violate the entropy 
inequalities: $v(E_+) = V(E_+,E_-)$. This line intersects  
the graph of $1/v(E)$ at point $(E_F, 1/I_F)$  on the first branch. 
It again intersects the graph at the common point of the
second and third branches 
$(E_m,1/v_m)$. Observe that on the part of the first branch 
that belongs to the domain, we have $v(E_-) = V(E_+,E_-)$, 
while on the second branch we have $E_+ = E_-$.
Also observe that $E_- = E^{(1)}(I_F)$, and $E_+ = E^{(3)}(I_F)$.

According to both our numerical experiments and asymptotic calculation,
the shock wave moves so that the point $(E_-,1/V) $ always 
remains on the boundary of the domain of ${\cal F}(E_-,u)$.
The shock wave first appears at a point on the second 
branch of $1/v(E)$ where  $E_+ = E_-$ (see Figure 4). Then the point 
$(E_-,1/V)$ moves from the second to the
first branch of $1/v(E)$, where it satisfies
$V(E_+,E_-)=v(E_-)$ until it arrives at $(E_F,1/I_F)$. It then continues 
moving on the upper boundary of the domain of 
${\cal F}(E_-,u)$ where $V(E_+,E_-)=v(E_+)$, until it reaches $(E_m,1/v_m)$ 
where the shock wave dies. (We have assumed a large enough bias, $\phi>\phi_d$.
For smaller biases, the shock reaches $x=L$ before the current has attained
its minimum value. Then the shock dies at some point of the upper boundary
of the domain of ${\cal F}(E_-,u)$).

 To summarize, in a full cycle of the oscillation the
following regions can be identified:
\begin{itemize}
\item Region I: One monopole 
(two domains) exists and the current is increasing
(see Figure 3.e).
\item Region II: A second monopole is born near $x=0$ and the 
current is nearing a maximum which is lightly higher than the peak of
the $v$ curve (see Figure 3.f).  
\item Region III: Two monopoles (three domains) propagate 
and the current decreases (see Figure 3.b).
\item Region IV: The monopole on the right disappears and the current
is nearing a minimum which is slightly higher than the minimum of
the $v$ curve (see Figure 3.c). 
\end{itemize}

The different regions here described 
are analyzed asymptotically in section \ref{sec-asymptotics},
where we derive the outer and inner solutions and put the pieces 
together working and matching in sequence on regions I,II,III and IV. 
The 
technical work on regions II and IV is done in sections \ref{sec-birth} and 
\ref{sec-death} respectively.

\setcounter{equation}{0}
\section{Kinematics of Shock Waves }      
\label{sec-kinematics}

In this section we shall analyze the processes of shock formation (near the
boundary) and the existence and properties of monopole wavefronts which are
a shock wave and a tail to its left or right moving at the shock velocity.
These matters are fundamental ingredients of the asymptotics of later sections.

\subsection{Formation of Shock Waves }  
   The formation of shock waves on the infinite real line from arbitrary 
initial data and constant current have been considered before for the model
of the Gunn effect in semiconductors \cite{kni,murray,bonilla91}. The main
result is that for an initial positive field profile a shock wave develops
in finite time if $\partial E/\partial x >0$ is large enough. For the 
time-periodic solution we are interested in, the shock waves develop near
the boundary at $x=0$ for appropriate values of the current density. We
shall now analyze Region II and derive exact formulas 
for the time and position at which a new shock wave appears. 

Let us solve the hyperbolic equation (\ref{1-1}) with the boundary condition
(\ref{1-3}) by the method of characteristics. 
\begin{eqnarray}
\frac{d E(t;\tau)}{dt} = I - v\left( E(t;\tau)\right) ,\label{chE}\\
\frac{d x(t;\tau)}{dt} = v\left( E(t;\tau)\right) ,\label{chx}
\end{eqnarray}
with the boundary conditions:
\begin{eqnarray}
E(\tau;\tau) = E_0 (\tau) , \label{E0}\\
x(\tau;\tau) = 0. \label{x0}
\end{eqnarray}

We have parametrized the characteristic curves by the times at
which they issue from the boundary at $x=0$. The boundary field $E_0 (\tau)$ 
has to be chosen so that the boundary condition $\partial E(0,t)/\partial x = 
c$ holds. Clearly,
\begin{equation}
c = \frac{\partial E}{\partial x}(0,t) = \frac{\partial E}{\partial \tau}(\tau;
\tau)\, \frac{\partial \tau}{\partial x} \mid_{x=0}\, .\label{c=}
\end{equation}
We derive some relations that will be of use.
Taking a $\tau$-derivative of the solution of Eqs.\ (\ref{chx}) and (\ref{x0})
\begin{equation}
x(t;\tau) = \int_{\tau}^{t} v\left( E(s;\tau)\right) \, ds ,\label{solchx}
\end{equation}
and then setting $x=0$ (equivalently, $t=\tau$) in the result, we obtain 
\begin{equation}
\frac{\partial x}{\partial \tau}(\tau;\tau) = - v\left( E(\tau;\tau)\right) =
- v\left( E_0(\tau)\right)\, .\label{derx}
\end{equation}
Eqs.\ (\ref{c=}) and (\ref{derx}) yield
\begin{equation}
\frac{\partial E}{\partial \tau}(\tau;\tau) = - c\, v(E_0 (\tau)).\label{E_tau}
\end{equation}
An equation for $E_0$ is found by taking a $\tau$-derivative of the 
boundary condition (\ref{E0}) and substituting (\ref{E_tau}) in the result: 
\begin{eqnarray}
\frac{d E_{0}}{d \tau} =  \frac{\partial E}{\partial t}(\tau;\tau) + 
\frac{\partial E}{\partial \tau}(\tau;\tau) = I(\tau) - v(E_0 (\tau)) - c\, 
v(E_0 (\tau)) ,\nonumber
\end{eqnarray}
that is
\begin{equation}
\frac{d E_{0}(\tau)}{d\tau} + (c + 1)\, v\left( E_{0}(\tau)\right) =  
I(\tau) . \label{bdryfield}
\end{equation}
This equation may be obtained directly by substituting $c=\partial E/
\partial x$ into the hyperbolic equation (\ref{1-1}).
Finally, we can combine (\ref{chE}) and (\ref{chx}) into the equation
\begin{equation}
x(t;\tau) + E(t;\tau) = \int_{\tau}^t I(s)\, ds + E_0 (\tau).\label{B1}
\end{equation}
We can now derive a formula for the time at which a shock wave first
appears.
This happens when the function $\tau(x;t)$, obtained by solving 
$x(t;\tau) = x$ for fixed $t$, first becomes multivalued. At the time of 
shock formation,
the function $x(t;\tau)$, has an inflection point with zero slope in the 
$\tau$ variable and therefore satisfies 
$\partial x(t;\tau)/\partial\tau = 0$, and $\partial^2 x(t;\tau)/\partial\tau^2
 = 0$. These are two conditions to be used in determining the time and 
location of the new shock wave. 
\begin{equation}
\frac{\partial x(t;\tau)}{\partial \tau} = \frac{d E_{0}}{d \tau} - I(\tau) 
- \frac{\partial E(t;\tau)}{\partial \tau} 
= - (1+c)\, v(E_0(\tau)) - \frac{\partial E(t;\tau)}{\partial \tau} ,\label{B2}
\end{equation}
where (\ref{bdryfield}) has been used. We can find $\partial E(t;\tau)/\partial
\tau \equiv E_{\tau}(t;\tau)$ by 
differentiating (\ref{chE}) and (\ref{E0}) with respect to $\tau$, using 
(\ref{E_tau}), and solving the resulting linear problem:
\begin{eqnarray}
\frac{dE_{\tau}(t;\tau)}{dt} = - v'\left( E(t;\tau)\right)\, E_{\tau}(t;\tau),
\label{B3}\\
E_{\tau}(\tau;\tau) = - c\, v\left( E_0 (\tau)\right) .\label{B4}
\end{eqnarray}
The result may be inserted into (\ref{B2}) to obtain:
\begin{equation}
\frac{\partial x(t;\tau)}{\partial \tau} = v(E_0 (\tau))\, \left( c\,\exp
\left[- \int_{\tau}^{t} v'(E(s;\tau))\, ds\right] - 1 - c\right)\, .\label{B7}
\end{equation}
From Eq.~(\ref{B7}) we immediately obtain
\begin{eqnarray}
\frac{\partial^{2} x(t;\tau)}{\partial \tau^{2}} = \frac{v'(E_0 (\tau))}{v(E_0 
(\tau))}\, E'_0(\tau)\, \frac{\partial x(t;\tau)}{\partial \tau} +
 c\, v(E_0 (\tau))\,\exp\left[- \int_{\tau}^{t} v'(E(s;\tau))\, ds\right] 
\nonumber\\
\times \left(  v'(E_0 (\tau)) + c\, v(E_0 (\tau))\, \int_{\tau}^{t} 
v''(E(r,\tau))\,\exp\left[- \int_{\tau}^{r} v'(E(s;\tau))\, ds\right]\, 
dr\right).   \label{B7b}
\end{eqnarray}
When the expressions given by Eqs.~(\ref{B7}) and (\ref{B7b}) are set equal to
zero, their solutions $t = t_s$ and $\tau = \tau_s$ yield the time and position 
at which a shock wave first appear,
\begin{eqnarray}
\frac{c}{1+c} = \exp\left[\int_{\tau_{s}}^{t_{s}} v'\left( E(t;\tau_{s})
\right)\, dt \right] ,  \label{nsw1}\\
-\frac{v'(E_0 (\tau_{s}))}{c\, v(E_0 (\tau_{s}))} = \int_{\tau_{s}}^{t_{s}} 
v''(E(r,\tau_{s}))\, \exp\left[- \int_{\tau_{s}}^{r} v'(E(s;\tau_{s}))\, 
ds\right]\, dr ,\label{nsw3}\\
x_s = \int_{\tau_{s}}^{t_{s}} v\left( E(t;\tau_{s})\right)\, dt . \label{nsw2}
\end{eqnarray}
The times $t_s$ and $\tau_s$ in Eqs.~(\ref{nsw1}) and (\ref{nsw2}) are 
functions of $c$ and $L$ determined by the condition that $t_s$ is the
smallest time for which (\ref{nsw1}) and (\ref{nsw3}) are satisfied. Notice 
that (\ref{nsw1}) has no solution 
$t_s > \tau_s$ for $-1\leq c\leq 0$. This
explains why the time-periodic electric field profiles, which appear
in the numerical simulations with such boundary conditions do not
exhibit shocks \cite{bkmw}. In order to find $t_s ,\, x_s$ and $\tau_s$
from equations (\ref{nsw1})-(\ref{nsw2}),
we need to know $E(t;\tau)$ along the characteristics, which is obtained
by solving (\ref{chE}) when the current is known.
This analysis is carried out in Section \ref{pieces} using an
asymptotic approximation to the current density $I(t)$.

\subsection{Monopole Wavefronts}

We now construct the ILs refered to in the Introduction, which are
basic elements of our asymptotics.      
Assuming constant current, we construct shock waves with a
left or right tail moving rigidly at the shock velocity $V(E_+,E_-)$. We call 
a {\em monopole with a left (resp.\ right) tail} the 
internal layer composed of the 
shock wave plus the transition layer (the tail) to its left (resp.\
right). The monopoles will be used as building blocks in the asymptotic
description of the time-periodic solution. In the moving coordinate
$\chi = x - X(t)$, where $X(t)$ is the position of the shock at time
$t$ [so $dX/dt = V(E_+,E_-)$], we have
\begin{eqnarray}
[v(E) - V(E_+,E_-)]\,\frac{\partial E}{\partial\chi} = I - v(E).
\label{tail}
\end{eqnarray}
We shall distinguish two cases:

{\em Case 1:}  $I<V$. We construct a left tail by solving (\ref{tail}) in the region 
$\chi\leq 0$. As $\chi$ varies from 0 to $\chi=-\infty$, $E$ varies from 
$E(0-)=E_-$ to $E(-\infty)=E_L$. The quantity $v(E)$ varies from $v(E_-)$
to $v(E_L) = I$ [$E_L$ is necessarily a fixed point of (\ref{tail})]. Let us
prove that $v(E_-)=V(E_+,E_-)$. If $V<v(E_-)$, we have $v(E_L) = I<V<v(E_-)$,
and the function $v(E)-V$ has a zero between $E_L$ and $E_-$. Thus (\ref{tail})
does not have a solution that connects $E_L$ and $E_-$, and we must have
$v(E_-)\leq V(E_+,E_-)$. But then the entropy condition $V(E_+,E_-)\leq v(E_-)$
implies $V(E_+,E_-)=v(E_-)$. The behavior of the tail near $\chi = 0$ is 
described by 
\begin{eqnarray}
\frac{dE}{d\chi}\sim  \frac{I - V}{v'(E_-) (E-E_-)},\quad\quad
E(0) = E_-,\nonumber
\end{eqnarray}
which when solved yields
\begin{eqnarray}
E(\chi)\sim  E_- - \sqrt{\frac{2(V-I)}{v'(E_-)}}\, \sqrt{-\chi},\quad\quad
\chi\rightarrow 0-.\nonumber
\end{eqnarray}
With this choice of the minus sign in front of the square root, $E(\chi)$
is an increasing function, as indicated by (\ref{tail}). Since $V>I$,
$v'(E)>0$ and therefore $E_-$ is necessarily on the first branch of the 
$v$ curve.

The only solution of (\ref{tail}) for $\chi>0$ with $E(0+)=E_+$ that 
we have observed in the numerical simulations is the constant solution
$E(\chi)=E_+$, with $v(E_+)=I$. We have the following partial explanation
of this fact. Let $E(\chi)\rightarrow E_R$ as $\chi\rightarrow\infty$. 
Clearly $v(E_R)=I$. In a neighborhood of $E_R$ we can multiply (\ref{tail}) 
by $v'(E)\sim v'(E_R)$ and obtain
\begin{eqnarray}
\frac{d}{d\chi}(v-I)\sim   \frac{v'(E_{R})}{V-I}\, (v-I).\nonumber
\end{eqnarray}
As $\chi\rightarrow\infty$, the point $v=I$ is unstable if $v'(E_R)>0$ and
stable if $v'(E_R)<0$. The unstable case corresponds to fields on the
third branch of the $v(E)$ curve, and the only solution having 
$E\rightarrow E_R$ [hence $v(E)\rightarrow v(E_R)$] as $\chi\rightarrow\infty$
is $E\equiv E_R = E_+$ ($v\equiv I$). In the stable case, we cannot exclude
right tails based on the above arguments. We have excluded them because 
they do not appear in our numerical simulations.

{\em Case 2:}  $I>V$. This case is completely analogous to case 1. We
obtain monopole waves with a right tail. To summarize:
\begin{itemize}
\item When $I<V$, $v(E_-)=V(E_+,E_-) \equiv U(E_+)$, $I=v(E_+)$, the 
monopole moves with velocity $dX/dt = U(E_+)$, and it has a left tail, i.e., 
$E_L<E_- < E_+ = E_R$. 
\item When $I>V$, $v(E_+)=V(E_+,E_-)\equiv W(E_-)$, $I=v(E_-)$, the monopole 
moves with velocity $dX/dt = W(E_-)$, and it has a right tail, i.e., $E_L=E_- < 
E_+ < E_R$.
\end{itemize}

The functions $U(E_+)$ and $W(E_-)$ are defined by the above relations.

 In Figure 1a a monopole with a right tail can be
observed in the field profile corresponding to time (2). This is better
displayed when plotting $\partial E/ \partial x$ as shown in the inset
of Figure 1b. Time (1) corresponds to a value of the 
current close to $I_F$ ($I_F=0.58$) where neither left nor right
tail exists.

\setcounter{equation}{0}
\section{Asymptotic Analysis of the Time-Periodic Solution}
\label{sec-asymptotics}

Let us describe one period of the oscillations for bias $1 < \phi <
E^{(3)}(1)$ (recall that $E^{(1)}(I) < E^{(2)}(I) < E^{(3)}(I)$
are the three roots of $v(E) - I$ for $v_m < I < 1$). 
It is convenient to redefine the time and space scales in
such a way that the SL length becomes 1:
\begin{eqnarray}
\epsilon = \frac{1}{L}\, ,\quad\quad y = \frac{x}{L}\, ,\quad\quad
s = \frac{t}{L}\, .\label{3-1}
\end{eqnarray}
Then Eqs.\ (\ref{1-1}) - (\ref{1-3}) become
\begin{eqnarray}
\frac{\partial E}{\partial s} + v(E)\, \frac{\partial E}{\partial y}
 &=& \frac{I - v(E)}{\epsilon},  \label{3-2}  \\
\int_{0}^{1} E(y,s)\, dy &=& \phi.           \label{3-3} \\
\frac{\partial E(0,s)}{\partial y} &=& cL. \label{3-4}
\end{eqnarray}
We shall describe the time-periodic solution of these
equations in the limit $\epsilon\rightarrow 0$ ($L\rightarrow\infty$)
by leading-order matched asymptotic expansions. Observe that $L$ is the 
nondimensional length of the superlattice [see Appendix \ref{sec-discrete},
equations (\ref{1.16}) and (\ref{a5})],
and therefore we analyze the limit in which the SL length is large
compared to $\epsilon\,\tilde{E}_{M} / (\tilde{N}_{D}\, e\,)$.

\subsection{Outer solutions}
Clearly the leading order of the outer expansion of the solutions to 
Eqs.\ (\ref{3-2}) - (\ref{3-3}) yields:
\begin{eqnarray}
I - v(E) = 0.\label{3-5}
\end{eqnarray}
Then the outer electric field is a piecewise constant function whose
profile is a succession of zeroes of $I - v(E)$, $E^{(k)}(I)$ ($k=1,2,3$),
separated by discontinuities. These discontinuities are the shock waves 
corresponding to monopoles with left or right tails moving with speeds given 
by (\ref{1-4}), as discussed in Section \ref{sec-kinematics}. Let us
suppose that $y=Y(s)$ represents the location of a shock wave separating two 
different solutions of (\ref{3-5}). According to Section \ref{sec-kinematics},
\begin{eqnarray}
\frac{dY}{ds} = U(E_+), \quad V(E_+,E_-) = v(E_-)\equiv U(E_+)\quad\mbox{if}
\quad I<V(E_+,E_-), \label{3-6}  \\
\frac{dY}{ds} = W(E_-), \quad V(E_+,E_-) = v(E_+)\equiv W(E_-)\quad\mbox{if}
\quad I>V(E_+,E_-),
 \label{3-7}
\end{eqnarray}
Notice that the limiting values $E_+$ in (\ref{3-6}) and $E_-$ in (\ref{3-7})
are solutions of (\ref{3-5}), and therefore they are functions of $I(s)$.
The current density should then be determined from these equations and the bias 
condition (\ref{3-3}), (see below).

\subsection{Inner solutions}
Near $y=0$ or near the shock waves there are regions of fast variations of
the electric field. In them $I=I(s)$, $E\sim F(x,t)$, where $x=y/\epsilon$
and $t=s/\epsilon$ according to (\ref{3-1}). 

The field in the boundary layer near $y=0$ obeys the equations:
\begin{eqnarray}
\frac{\partial F}{\partial t} + v(F)\, \frac{\partial F}{\partial x}
= I(s) - v(F),  \label{3-8}  \\
F(0,t) = E_0(s),\quad\mbox{with}\quad
\epsilon \frac{d E_0}{ds} + (1+c)\, v(E_0) = I(s), \label{3-9}
\end{eqnarray}
because of (\ref{bdryfield}). Except in very short time intervals when a new
shock wave is being formed, $F=F(x)$ is a quasi-stationary monotonically 
increasing profile joining $F(0) = E^{(1)}[I(s)/(1+c)]$ and $F(\infty) = 
E^{(1)}[I(s)]$.

In the tail regions of a monopole, the electric field 
is a solution of (\ref{tail}), that 
is, $F = F(\chi;s)$ with $\chi = (y-Y(s))/\epsilon$ and
\begin{eqnarray}
[v(F) - V(E_+,E_-)]\,\frac{\partial F}{\partial\chi} = I(s) - v(F).
\label{3-10}
\end{eqnarray}
When $I(s) < V(E_+,E_-)$, [resp.\ $I(s) > V(E_+,E_-)$], $E_-$ (resp.\ $E_+$) 
is a function of $E_+$ (resp.\ $E_-$) given by $V(E_+,E_-) = v(E_-)
\equiv U(E_+)$ [resp.\ $V(E_+,E_-) = v(E_+)\equiv W(E_-)$], and (\ref{3-10}) 
should be solved for $\chi<0$ (resp.\ $\chi>0$) with the boundary condition 
$F(0-) = E_-$ [resp.\ $F(0+) = E_+$]. Obviously $F$ matches the value of the
outer solution, $E^{(k)}[I(s)]$, as we leave the tail region, $\chi\rightarrow 
-\infty$ ($ k=1,2 $) or $\chi\rightarrow +\infty$ ($ k=2,3 $). 

\subsection{Putting the pieces together}
\label{pieces}

We shall start at a time where there is only one shock wave 
on the SL, the current is $I_F$ (see Fig.\ 4) and the electric 
field is a shock wave joining
spatially uniform regions with $E_-=E^{(1)}(I_F)$, $E_+=E^{(3)}(I_F)$. At time 
$s=0$, the shock wave is located at $y=Y_F = [E^{(3)}(I_F)-\phi]/[E^{(3)}(I_F) -
E^{(1)}(I_F)]$ according to the bias condition (\ref{3-3}). For $s>0$, the 
shock $(E_-,1/V)$ is at the upper boundary of the domain of 
$E_+ = {\cal F}(E_-,u)$ (see Introduction), so that $v(E_+) = 
V(E_+,E_-)\equiv W(E_-)$, $I<1$ and a monopole with a rigidly moving right tail 
is formed. Let $y=Y(s)$ be the shock position. Then the outer field profile is 
\begin{eqnarray}
E(y,s) = E^{(1)}\left(I(s)\right),\quad\quad\mbox{if}\quad\quad 0<y<Y(s),
\nonumber\\
E(y,s) = E^{(3)}\left(I(s)\right),\quad\quad\mbox{if}\quad\quad Y(s)<y<1.
\label{3-11}
\end{eqnarray}
The bias condition (\ref{3-3}) determines $Y(s)$ as a function of $I(s)$:
\begin{eqnarray}
Y = \frac{E^{(3)}(I)-\phi}{E^{(3)}(I) - E^{(1)}(I)}. \label{3-12}
\end{eqnarray}
Inserting (\ref{3-12}) into (\ref{3-7}) we find the following autonomous
equation for the current density:
\begin{equation}
\frac{dI}{ds} = \frac{(E^{(3)} - E^{(1)})^{2}\, v'_{1}\, W_{1}
}{ E^{(3)} - \phi + (\phi - E^{(1)})\, v'_{1}/v'_{3} } 
\geq 0\, ,	\label{3-13}
\end{equation} 
where $v'_j = v'(E^{(j)})$, and $W_j = W(E^{(j)})$, $j= 1,2,3$. To obtain
(\ref{3-13}) we have used
\begin{eqnarray}
\frac{dE^{(n)}(I)}{ds} = \frac{1}{v'(E^{(n)}(I))}\, \frac{dI}{ds},
\quad\quad n=1,3,\nonumber
\end{eqnarray}
which follows from time-differentiation of (\ref{3-5}) with $E=E^{(n)}(I)$. 
Eq.~(\ref{3-13}) explicitly displays the quasi-steady growth of the current 
during this stage of the oscillation. 
 The solution $I(s)$ of (\ref{3-13}) reaches 1 at a time 
$s=s_1$ which is numerically calculated. As we 
approach this time, $E^{(1)}(I)\sim 1 - [2\, (1-I)/|v''(1)|]^{1/2}$, and the
solution of (\ref{3-13}) becomes
\begin{eqnarray}
I \sim 1 - \frac{|v''(1)|}{2}\, [a (s_1 - s)]^2\, ,\label{I-asympt}\\
a = \frac{[E^{(3)}(1) - 1]^{2}\, W(1)}{E^{(3)}(1) - \phi}\, . \label{3-a}
\end{eqnarray} 
The corresponding outer electric field profile is
\begin{eqnarray}
E(y,s_1) = 1,\quad\quad\mbox{if}\quad\quad 0<y<Y(s_1),
\nonumber\\
E(y,s_1) = E^{(3)}(1),\quad\quad\mbox{if}\quad\quad Y(s_1)<y<1,
\label{3-14}
\end{eqnarray}
with
\begin{eqnarray}
Y(s_1) = \frac{E^{(3)}(1)-\phi}{E^{(3)}(1) - 1}. \label{ys1}
\end{eqnarray}
Obviously after this instant our approximations break down. What happens then? 

We shall see below that at $s=s_M$, $s_M - s_1 = O(\sqrt{\epsilon})$, $I$ 
reaches a maximum and then decreases, while the field to the left of the 
monopole, $E_L = E_-$, increases linearly with time, and the field to the 
right of the monopole, $E_R = E^{(3)}(I)$, decreases. When $I$ surpasses 1, 
$E_L$ can no longer be approximated by $E^{(1)}(I)$. In a short time 
interval about
$s=s_M$, $E_L$ and $I$ are close to 1, and the difference $I-v(E_L)$ eventually
acquires a positive value of the same order as the time derivative $\epsilon
dE_L/ds = O(\epsilon)$. Then $I-1 = O(\epsilon)$, $E_L - 1 = O(\epsilon^{1/2})$,
which happens in a time scale $s-s_M = O(\epsilon^{1/2})$. 
We thus make the ansatz
\begin{eqnarray}
\hat{s} = \frac{s-s_{M}}{\epsilon^{1/2}}\, ,\label{s-hat}\\
I \sim 1 + \epsilon\, \hat{I}(\hat{s}), \label{I-hat}\\
E_L - 1 \sim \epsilon^{1/2} \hat{E}_L(\hat{s}) ,\quad\quad
E_R - E^{(3)}(1) \sim \epsilon \hat{E}_R(\hat{s})\, . \label{hats}
\end{eqnarray}
In the time scale (\ref{s-hat}), the shock speed is $O(\epsilon^{1/2})$, 
\begin{eqnarray}
\frac{dY}{d\hat{s}} = \epsilon^{1/2}\, W(1) + O(\epsilon)\, ,  \label{3-15}
\end{eqnarray}
so that 
\begin{eqnarray}
Y(\hat{s}) = Y(s_M) + \epsilon^{1/2}\, W(1)\, \hat{s} + O(\epsilon).
\label{3-16}
\end{eqnarray}
[Notice that $Y(s_M)-Y(s_1) = O(\epsilon)$]. Inserting the field profile 
(\ref{hats}) and the time scale (\ref{s-hat}) into Eq.~(\ref{3-2}), we get
\begin{eqnarray}
\frac{d \hat{E}_{L}}{d\hat{s}} = \hat{I} - \frac{1}{2}\, v''(1)\,
\hat{E}_{L}^{2} ,  \label{A1}\\
\epsilon^{1/2}\,\frac{d \hat{E}_{R}}{d\hat{s}} = \hat{I} - 
v'\left(E^{(3)}(1)\right)\,\hat{E}_{R} .    \label{A2}
\end{eqnarray}
We therefore have
\begin{equation} 
\hat{E}_{R} = \frac{\hat{I}}{v'\left(E^{(3)}(1)\right)}\, ,\label{A3}
\end{equation} 
to leading order. Inserting the electric field profile and (\ref{3-16}) 
into the bias condition
(\ref{3-3}), we find $\hat{E}_{L}$ by equating terms of order $\epsilon^{1/2}$:
\begin{eqnarray}
\hat{E}_{L} = \frac{(E^{(3)}(1)-1)\,W(1)}{Y(s_1)} =  a\, \hat{s},
\label{A5}
\end{eqnarray}
where $a$ is given by (\ref{3-a}). To find $\hat I$, we substitute this result 
into (\ref{A1}), thereby getting
\begin{equation} 
\hat{I} = a - \frac{|v''(1)|}{2}\, a^{2}\, \hat{s}^{2}\, .\label{A4}
\end{equation} 
In outer units we therefore have
\begin{eqnarray} 
I \sim 1 + \epsilon a\, \left[ 1 - \frac{|v''(1)|\, a\, (s-s_{M})^{2}}{2
\epsilon}\right]\, ,\label{A66}\\
E_{L} \sim  1 + a\, (s-s_M). \label{A77}
\end{eqnarray} 

We have chosen $E_-(s_M) = 1$, as for this value of the electric field the 
current density reaches its maximum. The relation between $s_M$ and previous
times for which the approximations (\ref{3-11})-(\ref{3-13}) hold should be
found from a matching condition. 
$I(s)$ takes on the value $1-\kappa\epsilon$, $\kappa = O(1)$, at the times 
$s_{\kappa 1}$ and $s_{\kappa 2}$, given by
\begin{eqnarray}
s_{\kappa n} = s_{M} + (-1)^n \, \sqrt{\frac{2(a+\kappa)\epsilon}{a^{2}
|v''(1)|}}\, , \label{t'_M}
\end{eqnarray}
with $n=1,2$. $s_1$ for which $I(s_1)=1$ corresponds to $\kappa = 0$ in 
(\ref{t'_M}). At these times, $E_R \sim E^{(3)}(1)$ and 
\begin{eqnarray}
E_L(s_{\kappa n}) \sim 1 + (-1)^n \,\sqrt{\frac{2(a+\kappa)\epsilon}{|v''(1)|}}\, . 
\label{E-t'_M}
\end{eqnarray}
Notice that (\ref{A66}) and (\ref{I-asympt}) match for any positive $\kappa = 
O(1)$, as their difference is:
\begin{equation} 
\sqrt{2\, |v''(1)|\, a^{2}\, (a+\kappa)\, \epsilon}\,\, (s-s_{1}) + O(\epsilon)\, ,
\label{difference}
\end{equation} 
which is $O(\epsilon^{1/2}) = o(1)$ when the proper outer scales of current 
[$I = O(1)$] and time [$s=O(L)$] are used. We will choose appropriately the
value of $\kappa$ ($\kappa = 1.78$), so as to 
 optimize the leading-order asymptotics we are describing here.

During a time interval about $s=s_M$, the time derivative of $E_L$ 
is not negligible, and we have an unsteady stage during which a 
new shock wave is created if the excess charge $c$ is positive. 
Using the current density given by equation (\ref{A66}), the new 
shock wave appears at $x=x_s$, $t=t_s$ ($s_s\, , y_s$ in the outer 
variables) as obtained by solving Eqs.~(\ref{nsw1}) - (\ref{nsw2}).
Details about the birth of the new wave are given in Section 
\ref{sec-birth}.

After the new wave has appeared, there is a time interval where both 
old and new shocks (located at $y=Y$ and $y=Y_n$, respectively) coexist. 
The outer field profile consists of two shock 
waves connecting regions where the electric field is uniform:
\begin{eqnarray}
E = E_L,\quad\quad 0<y<Y_n\, ,\nonumber\\
E = E_M,\quad\quad Y_n<y<Y\, ,\nonumber\\
E = E_R,\quad\quad Y<y<L\, .\label{3-17}
\end{eqnarray} 
The bias condition (\ref{3-3}) is now
\begin{eqnarray}
Y_n\, E_L + (Y - Y_{n})\, E_{M} + (1 - Y)\, E_R = \phi .   \label{3-18}
\end{eqnarray}
Eq.~(\ref{3-5}) implies that 
\begin{equation}
E_L = E^{(1)}(I(s)),\quad\quad 
E_M = E^{(2)}(I(s)),\quad\quad 
E_R = E^{(3)}(I(s)). \label{3-19}
\end{equation} 
The equations of motion for the monopoles are
\begin{eqnarray}
\frac{dY_{n}}{ds} =  U(E_M), \label{3-20}\\ 
\frac{dY}{ds} = W(E_M) . \label{3-21} 
\end{eqnarray}

From Eqs.~(\ref{3-18}) - (\ref{3-21}), we determine the unknowns $I(s)$,
$Y_n(s)$ and $Y(s)$ as follows. The bias condition (\ref{3-18}) may be 
rewritten as
\begin{eqnarray} 
Y_{n} = \beta - \alpha\, Y, \label{3-22}
\end{eqnarray}
where
\begin{eqnarray} 
\alpha = \frac{E_{R} - E_{M} }{E_{M} - E_{L}}\, ,\quad\quad
\beta = \frac{E_{R} - \phi}{E_{M} - E_{L}}\, ,   \label{3-23}
\end{eqnarray}
or using (\ref{3-19}),
\begin{eqnarray} 
\alpha = \frac{E^{(3)} - E^{(2)} }{E^{(2)} - E^{(1)}}\, ,\quad\quad
\beta = \frac{E^{(3)} - \phi}{E^{(2)} - E^{(1)}}\, .   \label{3-23bis}
\end{eqnarray}
The three remaining unknowns $Y$, $Y_n$ and $I$ can be determined from Eqs.\
(\ref{3-19}) - (\ref{3-23}). By using these equations, we can first find
$Y$ and $Y_n$ as functions of $I$, and then obtain an equation for $I(s)$. 
Eqs.\ (\ref{3-20}) and (\ref{3-21}) yield
\begin{eqnarray} 
\frac{dY_{n}}{dI} = \frac{U_{2}}{W_{2}}\, \frac{dY}{dI}\, ,  \label{3-24}
\end{eqnarray}
A linear equation for $Y(I)$ may be obtained from (\ref{3-22}), (\ref{3-23bis}) 
and (\ref{3-24}):
\begin{eqnarray}
\frac{dY}{dI} + \frac{\alpha '\, W_{2}}{U_{2}+\alpha W_{2}} \, Y = 
\frac{\beta '\, W_{2}}{U_{2}+\alpha W_{2}}\, , \label{eqY} 
\end{eqnarray}
 where $\alpha '$ and $\beta '$ are the derivatives of
$\alpha$ and $\beta$ with respect to $I$.

A careful computation shows that the solution obeying $Y(1) = [E^{(3)}(1) - 
\phi]/[E^{(3)}(1) - 1]$ (leading-order location of the old shock wave at the 
time the new shock forms), and thus matches the previous stage, is
\begin{eqnarray} 
Y = \lim_{\gamma\rightarrow 0+}\, 
\left\{ \,
\int_{1-\gamma}^{I}\frac{W_{2}\,
\beta '}{U_{2}+\alpha W_{2}}\,\exp\left[-\int_{J}^{I}\frac{\alpha '\, 
W_{2}}{U_{2}+\alpha W_{2}}\, dr\right] \, dJ \nonumber
\right. \\
\left.
+\, \frac{E^{(3)}(1) - \phi}{E^{(3)}(1) - 1}\, \exp\left[-\int_{1-\gamma}^{I}
\frac{\alpha '\, W_{2}}{U_{2}+\alpha W_{2}}dr\right] 
\, \right\}.  \label{3-25}
\end{eqnarray}
Eq.\ (\ref{3-25}) holds as long as 
\begin{eqnarray} 
0 < Y\leq 1\quad\quad \mbox{and}\quad\quad v_m\leq I < 1. \label{condition}
\end{eqnarray}
Taking a time derivative of the 
bias condition and using (\ref{3-20}) and (\ref{3-21}), we find 
\begin{eqnarray} 
\frac{dI}{ds} =  \frac{U_{2} + \alpha\, W_{2}}{\beta ' - \alpha '\, Y}\, ,  
\label{3-26}
\end{eqnarray}
to be solved with the initial condition $I(s_s) = I_s$, (the value of the 
current at the shock formation time), which comes from matching
with the previous stage. 

The solution of this equation shows that the current decreases until 
one of the two conditions (\ref{condition}) breaks down. Either
(i) the shock wave at $Y(s)$ reaches $y=1$ at some time $s_d$, or (ii) the 
current $I(s)$ takes on the value $v_m$ corresponding to the minimum electron 
velocity at some time $s_m$ with $Y(s_m) < 1$. One possibility or the other is 
realized according to the value of the bias: Let $\phi_d$ be the bias for which 
$Y=1$ when $I=v_m$ in Eq.\ (\ref{3-25}). For $1\leq \phi<\phi_d$ we have 
possibility (i), whereas possibility (ii) is realized for $\phi>\phi_d$. 
In both cases, we are left after this stage with one 
monopole with left tail and $E_L = E^{(1)}(I)$, and $E_R = E^{(2)}(I)$ 
[case (i)], or $E_R = E^{(3)}(I)$ [case (ii)], moving 
toward $x=L$. The current density is determined from the 
following equation, analogous to (\ref{3-13}):
\begin{equation}
\frac{dI}{ds} = \frac{(E^{(k)} - E^{(1)})^{2}\, v'_{1}\, U_{k}
}{ E^{(k)} - \phi + (\phi - E^{(1)})\, v'_{1}/v'_{k} }\, ,	\label{d20}
\end{equation} 
where $k=2$ in case (i) or $k=3$ in case (ii). This equation has to be 
solved  for $s>s_d$ with the initial condition $I(s_d)$ in case (i), or for 
$s>s_m$ with initial data $I(s_m) = v_m$ in case (ii). 

In case (i), $I'(s_d) > 0$ if 
\begin{equation}
1 < \phi < \left. \frac{E^{(1)}\, v'_{1} + E^{(2)}\, |v'_{2}| }
{ v'_{1} + |v'_{2}| }\right|_{s=s_{d}} \,\equiv \phi_{\beta} .	\label{d21}
\end{equation} 
For the velocity curve we use, the bias interval (\ref{d21}) is very narrow,
and the corresponding values of $I(s_d)$ are very close to 1. We may have
a small-amplitude current oscillation corresponding to a supercritical
Hopf bifurcation. When the bias is larger, $I'(s_d) < 0$, and $I$ decreases 
until $I=v_m$ is reached and $E_R$ becomes $E^{(3)}(I)$. From there onwards, 
$I$ increases and the situation is the same as in case (ii) after the fast 
intermediate stage. We then have a current oscillation of amplitude
approximately given by $1-v_m$. The transition from small to large-amplitude
oscillation is extremely sharp, which accounts for the difficulty in observing
it in simulations or in real laboratory experiments. In both cases,
after some time the current density increases until the value $I_F$, the
monopole flips its tail to the right, and we have completed the asymptotic
description of one period of the self-sustained oscillation. Comparison
between our asymptotic solution and a numerical simulation is shown
in Fig.\ 5.

\setcounter{equation}{0}
\section{Birth of the new shock wave} 
\label{sec-birth}

In this Section we need to solve the characteristic equations given an
approximate expression for the current density, e.g. Eq.~(\ref{A66}),
and then calculate the shock formation time and its earliest position.
These quantities can be used as initial values for the following stage
of the oscillation as described in the previous Section.

There are two possibilities:
\begin{itemize}
\item $J(t_s)\approx 1$ at the shock formation time. Then the current density
may be approximated by Eq.~(\ref{A66}) during the process of shock formation. 
\item $J(t_s) < 1$, and Eq.~(\ref{A66}) ceases to hold during the last
part of the process of shock formation. 
After the current decreases below $I=1$, 
we again have a quasisteady stage with  
\begin{equation}
E_L \sim E^{(2)}(I),\quad\quad E_R \sim  E^{(3)}(I),\quad\quad  
Y \sim \frac{E^{(3)}(I) - \phi}{ E^{(3)}(I) - E^{(2)}(I) }.	\label{a30}
\end{equation} 
and 
\begin{equation}
\frac{dI}{ds} \sim \frac{(E^{(3)} - E^{(2)})^{2}\, v'_{2}\, 
W_{2}}{ E^{(3)} - \phi + (\phi - E^{(2)})\, v'_{2}/v'_{3} } \leq 0 ,	
\label{a31}
\end{equation} 
Clearly $I(s)$ and $E_R \sim E^{(3)}$ decrease while $E_- \sim E^{(2)}$
increases as the time elapses. Eq.~(\ref{a31}) approximates the current
density from the time it has decreased below 1 until the shock formation 
time. If no shock wave is formed as $E^{(2)}$ approaches the value $\phi$
(which occurs for $c$ sufficiently close to 0), 
$Y\sim 1$ according to (\ref{a30}), and the old shock wave exits at the 
receiving end of the SL. At this time the stationary uniform field profile 
$E=\phi$ is reached and maintained. These observations are in excellent 
agreement with the results of the numerical simulations. 
\end{itemize}

For a wide interval of $c$, Eq.~(\ref{A66}) describes the current density
during the full shock formation stage. We will see that the shock formation 
time decreases as $c$ increases for small $c$, it reaches a minimum and 
then increases rapidly with $c$. Actually, Eq.~(\ref{nsw1}) implies that
the integral of $v'(E(t;\tau_s))$ has to be negative, and therefore that
the field on the characteristic with $\tau=\tau_s$ needs to take values on
the second branch of the $v(E)$ curve. Typically the field $E_0(\tau)$
increases with $\tau$, it reaches a maximum, $E_0^{max}$,
at $\tau_{max}$ and then 
decreases to $E^{(1)}(I(\tau))$. We have to distinguish two cases
depending on whether $E_0^{max}$ surpasses 1, which happens for
enough small $c$. In fact, the critical value, $\tilde{c}$,
for which this maximum equals 1 occurs
when the current density is maximum
[observe that Eq.~(\ref{A66}) does not depend on $c$]. 
As $I_{max} \sim 1 + \epsilon a$,
then it follows from (\ref{bdryfield}) that $\tilde{c}=\epsilon a$.

Let us consider the case $c < \tilde{c}$ in which this maximum value 
surpasses 1.
Then the smallest time $t_s$ for which (\ref{nsw1}) is
satisfied,
is the one corresponding to the characteristic issuing from
$E_0(\tau_s)=1$.
For smaller $\tau_s$ there is initially a positive contribution to the
integral, leading to higher values of $t_s$. For larger $\tau_s$ 
the field along the characteristic quickly falls to $E^{(1)}(I)$
leading again to a positive contribution to the integral and therefore 
equation  (\ref{nsw1}) is not satisfied. 
For $E_m > E(t; \tau) > 1$ the integral of the characteristic
equation along the direction of increasing $t$ is unstable. Thus, we
carry out the integration of the characteristic in the direction of
decreasing $t$,
starting at $E^{(2)}(I(t_s))$ and ending at t such that
$E(t;\tau_s)=1$.
The value of $t_s$ for which equation  (\ref{nsw1}) is satisfied
gives the shock formation time and therefore the shock position.     
In this case,  
the shock formation time
decreases with increasing $c$, because the right hand side of (\ref{nsw1})
increases. We have solved this problem for $L=50, \, c=10^{-4}$ obtaining
$t_s - t_M = 7.08$, $x_s=7.66$ and $I(t_s)=0.93$. This value agrees
well with the result of direct numerical simulations of the model.

If $c > \tilde{c}$,
the smallest time $t_s$ for which (\ref{nsw1}) is
satisfied, is the one corresponding to the characteristic issuing from
$E_0^{max}$ ($\tau_s = \tau_{max}$).
The characteristic corresponding to this value, $\tau_s$,
starts by giving a positive contribution to the integral
until $E(t; \tau_s)$ reaches 1. Therefore as $c$ increases this 
positive contribution increases, and the shock formation time
increases, which has been verified by direct numerical simulation.

Let us now find approximations to Eqs.~(\ref{nsw1}) - 
(\ref{nsw2}) with the help of our previous work on intermediate scales
instead of integrating numerically the characteristic equations. Clearly 
we find $s_s - s_M = O(\epsilon^{1/2})$,  and $y_s = O(\epsilon^{1/2})$.
While the old shock wave corresponds to a monopole with a right tail moving 
with velocity (\ref{3-7}), the new shock wave corresponds to a monopole with 
a left tail, moving with velocity (\ref{3-6}) (see Figure 3b).

In order to estimate the shock formation time and position, we need to
calculate the solution of (\ref{nsw1}) and (\ref{nsw3}) approximately. This
might be quite involved, as the current density (needed to calculate the
electric fields appearing on the formulas) depends on the field profiles,
which should be calculated by the method of characteristics. Fortunately 
in the asymptotic limit $\epsilon\rightarrow 0$, the new shock wave is born 
during a time interval $|s-s_M| = O(\sqrt\epsilon)$, at a distance $y_s = 
O(\sqrt\epsilon)$ from $y=0$. The current density is approximately given
by the parabola (\ref{A66}), and both the equations for the field at the
injecting contact and the characteristic equations are simpler. We shall
give the approximate form of these equations and then insert their solutions
in the exact formulas (\ref{nsw1}) and (\ref{nsw3}), whose solution yields
the shock formation time.

Let us start finding the approximate equation for the electric field at $y=0$,
(\ref{bdryfield}). During the quasistationary first stage of the oscillation
described by Eqs.\ (\ref{3-11}) to (\ref{3-13}), the field at the SL boundary 
is given by the quasi-steady form of (\ref{bdryfield}),  
\begin{equation}
E_{0}(s)\sim E^{(1)}\left(\frac{I(s)}{1+c}\right)\, ,\label{in1}
\end{equation}
until $s=s_1$. During the faster stage about $s_M$ characterized by the scaling
(\ref{s-hat}) - (\ref{hats}), (\ref{bdryfield}) may be approximated by using
(\ref{A4}) for $\hat I$ and $E_0 - 1 = \epsilon^{1/2}\, \hat{E}_0$:
\begin{equation}
\frac{d\hat{E}_{0}}{d\hat{s}} = a - cL + \frac{|v''(1)|\, (1+c)}{2}\,
\left(\hat{E}_{0}^{2} - \frac{a^{2}\, \hat{s}^{2}}{1+c}
\right)\, . \label{in2}
\end{equation}
We can rewrite this equation so as to eliminate all parameters but one:
\begin{eqnarray}
\frac{de_{0}}{d\hat{\sigma}} = \mu  + e_{0}^{2} - \hat{\sigma}^{2}\, ,
\label{in4}\\
\mu = (1 - \frac{c}{a\epsilon})\, \sqrt{1+c}\, ,\label{in5}\\
e_{0} = \frac{|v''(1)|^{1/2}\, (1+c)^{3/4}}{\sqrt{2a}}\, \hat{E}_0\, ,
\label{in6}\\
\hat{\sigma} = \frac{(|v''(1)|\, a)^{1/2}\, (1+c)^{1/4}\,\hat{s}}{\sqrt{2}}\, .
\label{in7}
\end{eqnarray}
This equation should be solved with the initial condition 
\begin{equation}
E_0(s_{\kappa 1}) = E^{(1)}\left(\frac{1 - \kappa\epsilon}{1+c}\right)\, ,
\label{in3}
\end{equation}
which follows from (\ref{in1}) when the current $I(s_{\kappa 1}) = 1 - \kappa
\epsilon$ (for $s$ in the overlapping region) is used 
[see (\ref{t'_M})]. If $c=O(\epsilon)$, we 
may solve (\ref{in3}) to leading order within the scaling 
(\ref{in6})-(\ref{in7}):
\begin{equation}
e_0(- \sqrt{1+\kappa/a}) = - \sqrt{\frac{\kappa + c/\epsilon}{a}}\, .
\label{in3bis}
\end{equation}

A phase plane study of Eq.~(\ref{in4}) indicates that $E_0$ increases,
reaches a maximum and then decreases. There are several cases worth 
consideration according to the values of the parameter $\mu$ (Figure 6). 
If $c < a \epsilon$, $E_0$ surpasses 1 
(Fig.\ 6a). In this case we need to solve (\ref{in4}) in order to find 
the initial condition for the characteristic equations (\ref{in4}) with 
$\mu=1$. For smaller $\mu$'s, we may still use (\ref{in1}) to 
approximate $E_0$. In either case the shock formation time will be obtained 
by solving the characteristic equations with an initial condition given by
the solution of Eq.~(\ref{in4}) or by (\ref{in1}). 

We now find the approximate form of the characteristic equations. In the 
scaling (\ref{s-hat}) - (\ref{hats}) they become:
\begin{equation}
\frac{d\hat{E}}{d\hat{s}} = a + \frac{|v''(1)|}{2}\,
(\hat{E}^{2} - a^{2}\, \hat{s}^{2})\, . \label{ch1}
\end{equation}
By redefining the variables in this equation, we can rewrite it as
\begin{eqnarray}
\frac{de_{1}}{d\sigma} = 1  + e_{1}^{2} - \sigma^{2}\, ,\label{ch2}\\
e_{1} = \sqrt{\frac{|v''(1)|}{2a}}\, \hat{E}_0\, , \label{ch3}\\
\sigma = \sqrt{\frac{a |v''(1)|}{2}}\,\hat{s}\, . \label{ch4}
\end{eqnarray}
Notice that $\sigma = (1+c)^{ - 1/4}\,\hat{\sigma}$ by (\ref{in7}). When 
$\sigma$ takes on this value, the field on the characteristic is equal to the 
boundary field $E_0$. In scaled variables, the characteristic is parametrized
by the value $\hat{\sigma}$ such that when $\sigma = (1+c)^{ - 1/4}\,
\hat{\sigma}$, $y=0$ and 
\begin{eqnarray}
e_{1}\left(\frac{\hat{\sigma}}{(1+c)^{1/4}};\frac{\hat{\sigma}}{(1+c)^{1/4}}
\right) = (1+c)^{- 3/4}\, e_{0}(\hat{\sigma}).\label{in9}
\end{eqnarray}

The general solution of (\ref{ch2}) is
\begin{eqnarray}
e_{1}(\sigma;\sigma_{\tau}) = \sigma - \frac{1}{\int_{\sigma_{0}}^{\sigma} 
\exp(-\sigma^{2} + s^{2})\, ds}\, ,\label{in8}
\end{eqnarray}
for $\sigma > \sigma_0$ ($e_1\rightarrow - \infty$ as $\sigma\rightarrow
\sigma_0 +$), where $\sigma_{0}$ is a function of $\sigma_{\tau} = \hat{\sigma
}/(1+c)^{1/4}$ coming from the initial condition (\ref{in9}).

Inserting (\ref{in8}) in (\ref{nsw1}) and (\ref{nsw3}), we find
\begin{eqnarray}
\frac{1+c}{2}\,\ln(1+\frac{1}{c}) = \frac{\sigma_{s}^{2} - \sigma_{\tau}^{2}}{2}
+ \ln\frac{ \int_{\sigma_{0}}^{\sigma_{s}} \exp(s^{2}-\sigma_{s}^{2})\, ds}{ 
\int_{\sigma_{0}}^{\sigma_{\tau}} \exp(s^{2}-\sigma_{\tau}^{2})\, ds}\, ,
\label{in10}\\
\frac{a\epsilon}{c\sqrt{1+c}}\,\left( - \sigma_{\tau} + \frac{1}{\int_{\sigma_{0}}^{
\sigma_{\tau}} \exp(-\sigma_{\tau}^{2} + s^{2})\, ds}\right) = \nonumber\\
\int_{\sigma_{\tau}}^{\sigma_{s}} \left[ e^{(\sigma_{\tau}^{2}
- s^{2})/2}\, \frac{ \int_{\sigma_{0}}^{s} e^{r^{2}} \, dr}{ 
\int_{\sigma_{0}}^{\sigma_{\tau}} e^{r^{2}} \, dr}\right]^{2/(1+c)}\, ds\, .
\label{in101}
\end{eqnarray}
Here $\sigma_{s}$ and $\sigma_{\tau}$ correspond to $t_s$ and $\tau_s$ in 
Eq.~(\ref{nsw1}) - (\ref{nsw3}), respectively. Notice that $\sigma_{0}$ is a 
function of $\sigma_{\tau} = \hat{\sigma}/(1+c)^{1/4}$ given by (\ref{in9}).
We have solved numerically these equations for a fixed $\epsilon$ and
different values of $c$, obtaining a result consistent with that explained
earlier: the shock formation time decreases as $c$ increases for small $c$, 
it reaches a minimum and then increases rapidly with $c$.                                                                                                                              

\setcounter{equation}{0}
\section{Death of the monopole}
\label{sec-death}

We only need to consider case (ii) separately, for case (i) does not require
the introduction of different scales and was already analyzed in Section 
\ref{sec-asymptotics}. In case (ii), an intermediate stage with the same time 
scales as those for the shock birth describes the death of the old shock wave. 
We make the ansatz
\begin{eqnarray}
I \sim v_m + \epsilon\, \hat{I},\label{d1}\\
 E_{M,R} - E_m \sim \epsilon^{1/2} \hat{E}_{M,R} ,\label{d2}\\
 E_L - E_{m}^{(1)} \sim \epsilon \hat{E}_L ,\quad\quad
E_{m}^{(1)} = E^{(1)}(v_m), \label{d3}\\
\hat{s} = \frac{s-s_{m}}{\epsilon^{1/2}}\, ,\label{d4}
\end{eqnarray}
where $s_m$ is the time at which the current reaches its minimum value.
Equations (\ref{3-2}), (\ref{3-20}), and (\ref{3-21}) become
\begin{eqnarray}
\frac{dY_{n}}{d\hat{s}} = \epsilon^{1/2}\, U_{m} + O(\epsilon), \label{3-27}\\ 
\frac{dY}{d\hat{s}} = \epsilon^{1/2}\, v_m + O(\epsilon), \label{3-28}\\ 
\frac{d\hat{E}_{M}}{d\hat{s}} = \hat{I} - \frac{v''_{m}}{2}\, 
\hat{E}_{M}^{2} + O(\epsilon^{1/2}), \label{3-29}\\
\frac{d\hat{E}_{R}}{d\hat{s}} = \hat{I} - \frac{v''_{m}}{2}\, 
\hat{E}_{R}^{2} + O(\epsilon^{1/2}), \label{3-30}\\
\hat{I} - v'(E_{m}^{(1)})\, \hat{E}_{L} = O(\epsilon^{1/2}), \label{3-31}
\end{eqnarray}
where $U_m = U(E_m)$. Eqs.~(\ref{3-22}) and (\ref{3-23}) still hold, but
the approximation (\ref{3-23bis}) does not. The solution of (\ref{3-27}) is
\begin{eqnarray}
Y_n(\hat{s}) = Y_n(s_m) + \epsilon^{1/2}\, U_m\, \hat{s} + O(\epsilon).
\label{3-32}
\end{eqnarray}
We now insert (\ref{3-17}), (\ref{d2}), (\ref{d3}), and (\ref{3-32}) into
(\ref{3-22}) and (\ref{3-23}). The result is
\begin{eqnarray}
Y_n(s_m) + \epsilon^{1/2}\, U_m\, \hat{s} \sim \frac{E_{m} - \phi}{E_{m} 
- E_{m}^{(1)}} + \nonumber\\
\epsilon^{1/2}\,\frac{[\hat{E}_{R}\delta - (\delta - 1)\,\hat{E}_{M}]\, (\phi
- E_{m}^{(1)})}{(E_{m} - E_{m}^{(1)})^{2}} \, ,\label{3-33}
\end{eqnarray}
where we have defined
\begin{eqnarray}
\delta = \frac{E_{m} - E_{m}^{(1)}}{\phi - E_{m}^{(1)}}\, (1 - Y_{d})
\, . \label{3-34}
\end{eqnarray}
Here $Y_d$ is the position of the old shock wave when the current reaches
$v_m$, given by Eq.\ (\ref{3-25}) with $I = v_m$. Notice that $\delta\geq 0$;
$\delta = 0$ implies that the monopole $Y$ arrives at $x=L$ exactly when
the current density becomes $v_m$. From the bias condition
(\ref{3-33}), we find the position of the new shock at $s=s_m$, and the relation
\begin{eqnarray}
b\hat{s} = \hat{E}_{R}\,\delta - (\delta - 1)\,\hat{E}_{M}\, . \label{3-35}\\
b = \frac{U_{m}\, (E_{m} - E_{m}^{(1)})^{2}}{\phi - E_{m}^{(1)}}\, .
\label{3-36}
\end{eqnarray}
Notice that $\hat{E}_{M}<0<\hat{E}_{R}$. Then (\ref{3-35}) implies that
$\delta < 1$, for otherwise we could not have $\hat{s}<0$.
We now need to solve Equations (\ref{3-29}), (\ref{3-30}) and (\ref{3-35}) for 
the unknowns $\hat{E}_{M}$, $\hat{E}_{R}$ and $\hat{I}$. Notice that (\ref{3-35}) 
also holds at the end of the previous quasistationary stage where two
monopoles coexist. There Eqs.\ (\ref{3-29}) and (\ref{3-30}) with zero
left side also hold, which gives us the following expression for the current
at the end of the previous quasistationary stage:
\begin{eqnarray}
\hat{I} = \frac{v''_{m}}{2}\, \left( \frac{b\hat{s}}{2\delta - 1}
\right)^{2}\, ,\label{dmatch}
\end{eqnarray}
for $0\leq\delta <1/2$.

We now find the unknowns $\hat{E}_{M}$, $\hat{E}_{R}$ and $\hat{I}$. From 
(\ref{3-29}) and (\ref{3-35}) we can find $\hat I$ and  $\hat{E}_{M}$ 
as functions of $\hat{E}_{R}$, which inserted into (\ref{3-30}) yield 
\begin{eqnarray}
\frac{d\hat{E}_{R}}{d\hat{s}} = b - \frac{v''_{m}}{2(1-\delta)}\, (\hat{E}_{R}
- b\hat{s})\, [(1 - 2\delta)\, \hat{E}_{R} + b\hat{s}] , \label{3-37}
\end{eqnarray}
We can rewrite this equation so as to eliminate all parameters but $\delta$:
\begin{eqnarray}
\frac{df}{d\eta} = 1 - (f - \eta)\, [(1 - 2\delta)\, f + \eta] , \label{d14}\\
f = \left(\frac{v''_{m}}{2(1 - 2\delta)}\right)^{1/2}\,\hat{E}_M , \quad\quad 
\eta = \left(\frac{ v''_{m}}{2(1 - 2\delta)}\right)^{1/2}\, b\,\hat{s} , 
\label{d15}
\end{eqnarray}

A phase plane study of (\ref{d14}) indicates that all trajectories tend to
$f=\eta$ as $\eta\rightarrow\infty$, which matches with the quasistationary
stage that occurs after the disappearance of the old shock wave at $X_d$ (see
below). This is illustrated by Figure\ 6d and by the exact expression,
\begin{eqnarray}
f(\eta) = \eta + \frac{1}{(1 - 2\delta)\,\int_{\eta_{0}}^{\eta} 
\exp[(-\eta^{2} + s^{2})\,\delta]\, ds}\, ,\label{d16}
\end{eqnarray}
where $\eta_0$ is a constant. Of all these trajectories,the ones that
match the adiabatic stage before the old shock dies are those that
follow $f=-\eta/(1 - 2\delta)$ and experience a fast transition to
$f=\eta$. When $f=-\eta/(1 - 2\delta)$, the current matches (\ref{dmatch}).
Assume that the fast transition takes place at $\eta = \eta_1\gg 1$.
An approximate expression for $f(\eta)$ may be found letting $f = 
- \tilde{f}\eta_1/(1 - 2\delta)$, and $\eta = \eta_1 + \eta'$, with
$1\ll\eta'\ll\eta_1$. Inserting this ansatz in (\ref{d16}), we find
\begin{eqnarray}
\frac{d\tilde{f}}{d\eta'} =  - \eta_1\, (\tilde{f} + 1 - 2\delta)\, 
(1 - \tilde{f}) + O\left(\frac{1}{\eta_{1}}\right)\, , \label{d17}
\end{eqnarray}
whose solution yields 
\begin{eqnarray}
f(\eta) \approx - \frac{\eta_{1}}{1 - 2\delta}\,\left( 1 - \frac{2(1-\delta)}
{1+z\, e^{-2 (1-\delta)\eta_{1}\eta '}}\right)\, ,\label{d18}
\end{eqnarray}
where $z$ is a constant. After the fast transition ($\eta'\rightarrow\infty$), 
both $\hat{E}_M$ and $\hat{E}_R$ become $b\hat s$, the old shock dies and the 
current becomes
\begin{eqnarray}
I \sim v_m + \epsilon\, b + \frac{v''_{m}}{2}\, b^2\, (s-s_{m})^{2}\, .
\label{d19}
\end{eqnarray}

It is interesting to find out when the current $I$ reaches the value $v_m$.
In terms of the variables $\eta$ and $f$, $\hat I$ may be written as
\begin{eqnarray}
\hat I =  \frac{(\eta +f\delta)^{2} - 1 - \frac{df}{d\eta}\delta}{1-\delta}\, 
.\label{d201}
\end{eqnarray}
Notice that $\hat I = 0$ if $f=-\eta/\delta$. Then the current reaches $I=v_m$
($\hat I = 0$) if the slope of the line $f=-\eta/\delta$ is larger than the
slope of $f(\eta)$ as $\eta\rightarrow -\infty$, i.e. $1/\delta > 1/(1-2\delta)$.
This implies $0\leq\delta<1/3$, which occurs for biases $\phi_d<\phi<
\phi_{\delta}$. When the bias is larger, $1/3<\delta<1/2$, and the minimum of
the current is larger than $v_m$.

\setcounter{equation}{0}
\section{Concluding remarks}
\label{sec-discussion}

We have developed an asymptotic theory for the continuum limit of a
discrete drift model of doped SLs. Our theory is in excellent 
quantitative agreement with numerical simulations of the discrete model
for long enough SLs. Our asymptotic analysis of the continuum equations 
indicates that they have a stable time-periodic solution for an 
appropriate bias. That this is indeed so, remains an important open 
mathematical problem. The time-periodic oscillation is due to repeated
creation and propagation of traveling monopole wavefronts, formed by
a shock wave and an attached tail region moving at the same velocity as the
shock. The study of the shock-wave dynamics allows us to
reduce the complexity of the problem by decomposing it into ``coherent''
structures and analyzing their simpler reduced dynamics. This decomposition
should in principle yield a full explanation of the classical chaos
observed numerically when an appropriate harmonic forcing is added to
the bias \cite{BB}.

The present phenomena are related to the well-known Gunn instability in
bulk semiconductors \cite{shawold,murray,HB,bonilla91,onset}. The main
mathematical difference is that the traveling waves responsible for the
Gunn effect are solitary waves of the electric field (charge dipoles),
while in the present case they are monotone wavefronts of the electric
field (charge monopoles). We remark that the first numerical observation 
of a ``Gunn-like'' instability (in a drift-diffusion model with different
boundary conditions from ours) mediated by monopoles is due to H. Kroemer,
\cite{kro66}. An incomplete asymptotic study of the Kroemer drift-diffusion 
model valid for particular electron velocities was performed in \cite{HB}. 
The fact that many different models (e.g., ``slow'' Gunn effect in ultrapure 
Germanium \cite{BT,smsig,inma}) present the Gunn instability poses the 
problem of characterizing the model features that produce the Gunn effect, 
as precisely as possible.

Other open problems presently being considered include explaining how well
the asymptotic solution approximates the solution of the discrete equations
and how well these results compare with experiments \cite{grahnprl,merlin}. 
To this end, it is important to extend the asymptotic analysis to a more 
complete model of SLs under photoexcitation \cite{ICPF94,bonillaSL}. Finally,
a careful derivation of the discrete drift model from a fully quantum-mechanical
setting would be most desirable. None of the presently known derivations is
satisfactory.


\section*{Acknowledgments}
We thank J.\ Gal{\'a}n, H.\ T.\ Grahn, J.\ Kastrup, S.-H. Kwok,
R.\ Merlin and A.\ Wacker for fruitful discussions and collaboration 
on related topics.

\appendix 
\renewcommand{\theequation}{A.\arabic{equation}}
\setcounter{equation}{0}
\section{Brief Derivation of the Continuum Model 
from the Discrete Model} 
\label{sec-discrete}

A SL is a periodic 
succession of alternating long (for simplicity) slabs of two semiconductors 
(GaAs and AlAs in \cite{grahnprl,merlin}). 
Since the energy bandgap is different for the two semiconductors, the SL may be
thought of as a periodic succession of potential ``barriers'' and ``valleys''.
Along the growth direction of a finite SL, we find N periods formed by a
barrier and a valley with a typical length of tens of nanometers. The lateral
dimension may be one or ten thousand times larger, so that transport 
phenomena along the growth dimension may be considered to be one-dimensional.
In the simplest case of a n-doped N-period SL without laser illumination,
the discrete drift model consists of the following system of equations
\cite{bonillaSL,ICPF94}:

\begin{eqnarray}
\tilde{E}_{j} - \tilde{E}_{j-1} = \frac{e\,\tilde{l}}{\epsilon}\,
(\tilde{n}_{j} - \tilde{N}_{D}),          \label{poisson}\\
\epsilon\,\frac{d \tilde{E}_{j}}{d \tilde{t}} +
e\,\tilde{v}(\tilde{E}_{j})\,\tilde{n}_{j} = \tilde{J},
                                                   \label{ampere} \\
\frac{1}{N}\,\sum_{j=1}^{N} \tilde{E}_{j} = \frac{\tilde{\Phi}}{N\, \tilde{l}}.
\label{bias}
\end{eqnarray}
where $j= 1,\dots, N$. In this model Eq.\ (\ref{poisson}) and Eq.\
(\ref{ampere}) are, respectively, the one-dimensional Poisson equation 
(averaged over one SL period) and Amp{\`e}re's law for the electric field
$\tilde{E}_{j}(\tilde{t})$ and electron density $\tilde{n}_{j}(\tilde{t})$
at the site $j$, and for the total current density $\tilde{J}(\tilde{t})$. 
In these equations, the positive constants $\epsilon$, $\tilde{N}_{D}$, $l$ and
$e$ are the average permittivity, average donor concentration, SL period and
the absolute value of the charge of the electron. $\tilde{v}(\tilde{E})$
is an effective electron velocity to be specified later. Equation (\ref{bias}) 
establishes that the average electric field is given by the $dc$ voltage bias
$\tilde\Phi$. Notice that there are 2N+2 unknowns: $\tilde{E}_{0},\tilde{E}_{1},\ldots,
\tilde{E}_{N}, \tilde{n}_{1},\ldots,\tilde{n}_{N},\tilde{J}$ and 2N+1
equations so that we need to specify one boundary condition for $\tilde{E}_{0}$
plus an appropriate initial profile $\tilde{E}_{j}(0)$. The boundary condition
for $\tilde{E}_{0}$ (the average electric field {\em before} the SL) can be
fixed by specifying the electron density at the first site, $\tilde{n}_{1}$,
according to (\ref{poisson}). In typical experiments the region before the
SL has an excess of electrons due to a stronger n-doping there than in the
SL \cite{grahnprl,merlin}. Thus it is plausible assuming than there is an 
excess number of electrons at the first SL period measured by a dimensionless 
parameter $c$:
\begin{equation}
\tilde{E}_{1}(\tilde{t}) - \tilde{E}_{0}(\tilde{t}) = 
\frac{c\, e\,\tilde{l}\, \tilde{N}_{D}}{\epsilon}. \label{bc}
\end{equation}
$c$ has to be quite small because it is known that a steady 
uniform-electric-field profile is observed at low laser illumination in undoped 
SL \cite{merlin,ICPF94}. If we adopt (\ref{bc}) with $0<c\ll 1$ as our boundary 
condition, a steady almost uniform electric field profile is clearly a possible 
solution of Eqs.\ (\ref{poisson}) to (\ref{bias}).

The only function not specified so far is the electron velocity 
$\tilde{v}(\tilde{E})$. 
 This function is used to model sequential resonant 
tunneling (SRT) \cite{bonillaSL} in the SL, which is the 
main charge transport mechanism for the high electric fields.
The velocity curve can be derived from experiments \cite{bonillaSL}
or from simple one-dimensional quantum-mechanical calculations of 
resonant tunneling, as was done by Prengel et al \cite{scholl}.
For our analysis,
the only important characteristic of the velocity curve is that
it has to have a local maximum and a local minimum 
and therefore a region in which the velocity decreases with
increasing field. In our analysis we will use the curve 
depicted in the inset of Fig. 1a.

It is convenient to render the equations (\ref{poisson})-(\ref{bc}) 
dimensionless by adopting as the units of electric field and velocity the 
values at the first positive maximum of the velocity curve, $\tilde{v}
(\tilde{E})$, $\tilde{E}_{M}$ and $\tilde{v}_M$ (about $10^{5}$V/cm and 427 
cm/s, respectively for the sample of Ref.\ \cite{grahnprl}). We set
\begin{eqnarray}
E_j = \frac{\tilde{E}_{j}}{\tilde{E}_{M}}\, , \quad\quad n_j = 
\frac{\tilde{n}_{j}}{\tilde{N}_{D}}\, ,
\quad\quad I =  \frac{\tilde{J}}{e\tilde{N}_{D}\tilde{v}_{M}}\, ,
\nonumber\\
t = \frac{\tilde{v}_{M}\nu\,\tilde{t}}{\tilde{l}}\, , 
\quad\quad \phi = \frac{\tilde{\Phi}}{N\,\tilde{E}_{M}\,\tilde{l}}, 
\label{tilden}
\end{eqnarray}
where the dimensionless parameter $\nu$, defined as
\begin{equation}
\nu = \frac{\tilde{N}_{D}\, e\,\tilde{l}}{\epsilon\,\tilde{E}_{M}},
\label{1.16}		
\end{equation}
is about 0.1 for the SL used in the experiments \cite{grahnprl}. Then the 
dimensionless equations of the model are:
\begin{eqnarray}
E_{j} - E_{j-1} &=& \nu\, (n_{j} - 1),       \label{1}  \\
\frac{d E_{j}}{d t} + v(E_{j})\, n_{j} &=& I,  \label{2}  \\
\frac{1}{N}\,\sum_{j=1}^{N} E_{j} &=& \phi,           \label{3}  \\
E_{1} - E_{0} &=& \nu\, c; \quad\quad\quad  c \ll 1.				\label{4}
\end{eqnarray}
Here $v(E) = \tilde{v}(\tilde{E})/\tilde{v}_M$, and $\phi$ is a dimensionless 
control parameter (the $dc$ bias), whereas $\nu$, 
$c$ and $N$ are fixed for each SL. Equations (\ref{1}) and (\ref{2}) are to be 
solved with initial conditions for the fields, $E_{j}(0)$, compatible with 
the bias (\ref{3}) and the boundary condition (\ref{4}). The initial conditions 
for the electron density, $n_{j}(0)$, then follow from (\ref{1}).

In the continuum limit $\nu\rightarrow 0,\, N\rightarrow\infty$ so that
\begin{equation}
x = j \nu \in [0,L],\, \, L = N\nu\gg 1,                    \label{a5}  
\end{equation}
the system of equations (\ref{1})-(\ref{4}) becomes (\ref{1-1})-(\ref{1-3})
after the electron density $n(x,t)$ is eliminated by means of Poisson's 
equation. The hyperbolic equation (\ref{1-1}) develops shock waves from 
initial data in finite time \cite{bonilla91,murray}. To 
determine their velocity and entropy condition we have to go back to the 
discrete model (\ref{1})-(\ref{4}) \cite{ICPF94}. 
The shock wave is a field profile that moves rigidly with an average speed 
$V(E_+,E_- )$, so that $E_j (t) = E(j - j_s (t))$ with $E(-\infty) = E_-$, 
$E(+\infty) = E_+$  and $j_s (t + \Delta t) - j_s (t)\sim V(E_+,E_- )\,
\Delta t/\nu$, for large enough $\Delta t$. $j_s (t)$ is the QW where the 
profile $n_j (t)$ reaches its maximum value at time $t$ and  $X(t) =\nu\, 
j_s (t)$ corresponds to the shock position in the continuum limit.
Then $E_j (t + \Delta t) - E_j (t)$ is approximately given by
the distance $j_s (t + \Delta t) - j_s (t)$ that the shock
has advanced during $\Delta t$ times the field difference $- (E_j - E_{j-1})$
at some intermediate time in $(t,t + \Delta t)$. Thus we have
\begin{equation}
\frac{dE_{j}}{dt} \sim - \frac{E_{j} - E_{j-1}}{\nu}\, V(E_+,E_- ).
 \label{a10}
\end{equation}
If we now sum $(E_j - E_{j-1})$ over the shock region, we obtain
$E_+ - E_-$. On the other hand $n_j = O(\nu^{-1})\gg 1$ and $I = O(1)$ in
the shock region, so that (\ref{1}), (\ref{2}) and (\ref{a10}) yield
\begin{eqnarray}
E_+ - E_- = \sum_{j=-\infty}^{\infty} (E_j - E_{j-1}) \sim 
- \sum_{j=-\infty}^{\infty} \frac{\nu}{v(E_{j})}\,
\frac{dE_{j}}{dt}                 											\nonumber\\
 \sim V(E_+,E_- )\, \sum_{j=-\infty}^{\infty} 
\frac{E_{j} - E_{j-1}}{v(E_{j})}.															 \label{a11}
\end{eqnarray}
In the continuum limit this becomes the equal area rule (\ref{1-4})
\begin{eqnarray}
\int_{E_{-}}^{E_{+}} \left( \frac{1}{v(E)} - \frac{1}{V(E_{+},E_{-})} \right) 
\, dE = 0.										 \nonumber
\end{eqnarray}
To keep $n_j \geq 0$ inside the shock we must add the restrictions 
\begin{eqnarray}
v(E_- )\geq V(E_{+},E_{-})\geq v(E_+ ) \nonumber
\end{eqnarray}
(and hence $V=v(E)$ at only one point $x$ inside the shock). Likewise, no 
shock with $E_+ < E_-$ arises from realistic initial conditions with $n_j 
\geq 0$ \cite{HB}. Typical solutions of the equal area rule (\ref{1-4})
compatible with the entropy condition (\ref{1-5}) are depicted in  Fig.\
2. Notice that we can have admissible values of $E_-$ and $E_+$ 
belonging to equal or different branches of $1/v(E)$. However for $E_-$ and 
$E_+$ to be on the same branch of $1/v(E)$, they must belong to the second
branch, with $v'(E) < 0$.


\bigskip
\begin{center}
{\bf FIGURE CAPTIONS}
\end{center}

\noindent {\em Figure 1}. 
(a) Time evolution of the electric field profile on the SL 
using the velocity curve shown in the inset. 
(b) Charge density profiles, $\partial E(x,t)/\partial x$, showing the 
location of the IL for different times. The total current density versus 
time is shown in the leftmost inset, in which we have marked the times 
corresponding to the profiles depicted in part (a). The rightmost inset
shows clearly a monopole with a right tail.
 
\noindent {\em Figure 2}. Equal-area rule relating the electric field values
behind, $E_-$, and in front, $E_+$, of the shock wave with its velocity
$V(E_+,E_-)$.


\noindent {\em Figure 3}. The time-periodic solution in a nutshell.
(a) Motion of one shock during one period on a cylindrical surface
whose axis is space. The diameter of the cylinder basis is $1<1/I<1/v_m$.
(b) to (f) highlight of the equal-area conditions for the shock waves appearing
in the electric field profile. They correspond to the different stages 
marked in Figure 1b.

\noindent {\em Figure 4}. Domain of the function $E_+ = {\cal F}(E_-,1/V)$. The 
projection of $E_+(t)$ on this diagram for each time represents the evolution
of the shock wave during its entire life.

\noindent {\em Figure 5}. Comparison between our asymptotic solution and
a numerical simulation of the full model for $\phi = 1.25$, $\epsilon = 0.02$,
and $c=0.01$. 

\noindent {\em Figure 6}. Phase plane study of Eq.~(\ref{in4}) corresponding
to the field at $x=0$ during the stage of shock birth. (a) $\mu = 1$, which
corresponds to the electric field on the characteristics that emanate from
$x=0$. (b) $\mu = 0.5$. (c) $\mu = - 0.5$. (d) Phase plane describing the death
of a shock wave when $I\approx v_m$. It corresponds to
Eq.\ (\ref{d14})  with $\delta = 0.3$.


\begin{thebibliography}{99}
\bibitem{sserna}
J. G. BLOM, J. M. SANZ-SERNA and J. G. VERWER, {\em On simple moving grid 
methods for one-dimensional evolutionary partial differential equations}. 
J. Comput. Phys. {\bf 74} (1988), pp.~191-213.

\bibitem{bonilla91}
 L. L. BONILLA, {\em Solitary waves in semiconductors with finite geometry 
and the Gunn effect}. SIAM J. Appl. Math. {\bf 51} (1991), pp.~727-747. 

\bibitem {smsig}
L. L. BONILLA, {\em Small signal analysis of spontaneous current instabilities 
in extrinsic semiconductors with trapping: application to p-type ultrapure 
Germanium}. Phys. Rev. B {\bf 45} (1992), pp.~11642-11654.

\bibitem {ICPF94}
L. L. BONILLA, {\em Dynamics of electric field domains in superlattices}, in 
Nonlinear Dynamics and Pattern Formation in Semiconductors and Devices, 
edited by F.-J.\ Niedernostheide. Springer, Berlin, 1995, pp.~1-20.

\bibitem {inma-onset}
L. L. BONILLA I. R. CANTALAPIEDRA, M. J. BERGMANN and S. W. TEITSWORTH, {\em 
Onset of current oscillations in extrinsic semiconductors under dc voltage bias}.
 Semicond. Sci. Technol. {\bf 9} (1994), pp.~599-602.

\bibitem {bonillaSL}
L. L. BONILLA, J. GAL{\'A}N, J. CUESTA, F. C. MART{\'\i}NEZ and J. M. MOLERA,
{\em Dynamics of electric field domains and oscillations of the photocurrent 
in a simple superlattice model}. Phys. Rev. B {\bf 50} (1994), pp.~8644-8657.

\bibitem {onset}
L. L. BONILLA and F. J. HIGUERA, {\em The onset and end of the Gunn effect in
extrinsic semiconductors}. SIAM J. Appl. Math. {\bf 55} (1995), pp.~1625-1649. 

\bibitem{BT}
L. L. BONILLA and S. W. TEITSWORTH, {\em Theory of periodic and solitary space 
charge waves in extrinsic semiconductors}. Physica D {\bf 50} (1991), 
pp.~545-559.

\bibitem{BB}
O.\ M.\ BULASHENKO and L.\ L.\ BONILLA, {\em Chaos in resonant-tunneling 
superlattices}. Phys. Rev. B {\bf 52} (1995), pp.~7849-7852.

\bibitem {inma}
I. R. CANTALAPIEDRA L. L. BONILLA, M. J. BERGMANN and S. W. TEITSWORTH, 
{\em Solitary wave dynamics in extrinsic semiconductors under dc voltage 
bias}. Phys. Rev. B {\bf 48} (1993), pp.~12278-12281.

\bibitem {esaki}
L. ESAKI and L. L. CHANG, {\em New transport phenomenon in a semiconductor 
superlattice}. Phys. Rev. Lett. {\bf 33} (1974), pp.~495-498.

\bibitem {grahn90}
H. T. GRAHN, H. SCHNEIDER and K. VON KLITZING, {\em Optical studies of electric 
field domains in} GaAs-Al$_{x}$Ga$_{1-x}$As {\em superlattices}. Phys. Rev. B 
{\bf 41} (1990), pp.~2890-2899.

\bibitem {HB}
F. J. HIGUERA and L. L. BONILLA, {\em Gunn instability in finite samples of GaAs. 
II Oscillatory solutions in long samples}. Physica D {\bf 57} (1992), pp.~161-184.

\bibitem {kastrup}
J. KASTRUP, H. T. GRAHN, K. PLOOG, F. PRENGEL, A. WACKER and E. SCH{\"O}LL,
{\em Multistability of the current-voltage characteristics in doped GaAs-AlAs 
superlattices}. Appl. Phys. Lett. {\bf 65} (1994), pp.~1808-1810.

\bibitem{bkmw}
J. KASTRUP, H. T. GRAHN, R. HEY, K. PLOOG, L. L. BONILLA, M. KINDELAN, M. 
MOSCOSO, A. WACKER and J. GAL{\'A}N, {\em Electrically tunable GHz oscillations 
in doped GaAs-AlAs superlattices}. Phys.\ Rev.\ B {\bf 55}(1), (1997), to 
appear.

\bibitem {grahnprl} 
J. KASTRUP, R. KLANN, H. T. GRAHN, K. PLOOG, L. L. BONILLA, J. GAL{\'A}N, M. 
KINDELAN, M. MOSCOSO and R. MERLIN, {\em Self-oscillations of domains in 
doped GaAs-AlAs superlattices}. Phys. Rev. B {\bf 52} (1995), pp.~13761-13764.

\bibitem {kazar}
R. F. KAZARINOV and R. A. SURIS, {\em Electric and electromagnetic properties of 
semiconductors with a superlattice}.  Sov. Phys.--Semicond. {\bf 6}
(1972), pp.~120-131.

\bibitem {kni}
B.\ W.\ KNIGHT and G.\ A.\ PETERSON, {\em Nonlinear analysis of the Gunn effect}. 
Phys.\ Rev.\ {\bf 147} (1966), pp.~617-621.

\bibitem {kro66}
H. KROEMER, {\em Nonlinear space-charge domain dynamics in a semiconductor 
with negative differential mobility}. IEEE Trans. on Electron Devices {\bf 
ED-13} (1966), pp.~27-40.

\bibitem {merlin}
S.-H. KWOK, T. C. NORRIS, L. L. BONILLA, J. GAL{\'A}N, J. A. CUESTA, 
F. C. MART{\'\i}NEZ, J. M. MOLERA, H. T. GRAHN, K. PLOOG and R. MERLIN, 
{\em Domain wall kinetics and tunneling-induced instabilities in 
superlattices}.  Phys. Rev. B {\bf 51} (1995), pp.~10171-10174.

\bibitem {laik}
B. LAIKHTMAN, {\em Current-voltage instabilities in superlattices}. 
Phys. Rev. B {\bf 44} (1991), pp.~11260-11265.

\bibitem{lamiller}
B. LAIKHTMAN and D. MILLER, {\em Theory of current-voltage instabilities in 
superlattices}. Phys.\ Rev.\ B {\bf 48} (1993), pp.~5395-5412.

\bibitem{mila}
D. MILLER and B. LAIKHTMAN, {\em Theory of high field domain structrures in 
superlattices}. Phys.\ Rev.\ B {\bf 50} (1994), pp.~18426-18435.

\bibitem{murray}
J. D. MURRAY, {\em On the Gunn effect and other physical examples of perturbed 
conservation equations}. J.\ Fluid Mech.\ {\bf 44} (1970), pp.~315-346.

\bibitem {scholl}
F. PRENGEL, A. WACKER and E. SCH{\"O}LL, {\em A simple model for multistability 
and domain formation in semiconductor superlattices}. Phys. Rev. B {\bf 50} 
(1994), pp.~1705-1712.

\bibitem{shawold}
M. P. SHAW, H. L. GRUBIN and P. R. SOLOMON, {\it The Gunn-Hilsum
Effect}. Academic P., New York 1979.

\end{thebibliography}
\end{document}